\documentclass[singlecolumn,amsmath,amssymb,prb]{revtex4}
\usepackage{graphicx}
\usepackage{subfig}
\usepackage[countmax]{subfloat}
\usepackage{amsmath, amssymb}

\begin{document}


\title{The Effect of Disorder on Local Electron Temperature in Quantum Hall Systems}

\author{N. Boz Yurda\c{s}an}
\address{Dokuz Eyl\"{u}l University, Physics Department, Faculty of Sciences, 35160 Izmir, Turkey }
\author{S. Erden G\"{u}lebaglan}

\address{Y\"{u}z\"{u}nc\"{u} Y{\i}l University,Van Vocational School, Department of Electricity and Energy, 65080  Van,  Turkey }
\author{A. Siddiki}
\address{Maltepe University, Faculty Engineering and Natural Sciences, Department of Electrics and Electronics, 34857  Istanbul,  Turkey }

\begin{abstract}
{The local electron temperature distribution is calculated considering a two dimensional electron system in the integer quantum Hall regime in presence of disorder and uniform perpendicular magnetic fields. We solve thermal-hydrodynamical equations to obtain the spatial distribution of the local electron temperature in the linear-response regime. It is observed that, the variations of electron temperature exhibit an antisymmetry regarding the center of the sample in accordance with the location of incompressible strips. To understand the effect of sample mobility on the local electron temperature we impose a disorder potential calculated within the screening theory. Here, long range potential fluctuations are assumed to simulate cumulative disorder potential depending on the impurity atoms. We observe that the local electron temperature strongly depends on the number of impurities in narrow samples.}
\end{abstract}

\maketitle
\section{Introduction}

The quantized Hall effects (QHE)~\cite{Klitzing1980,Tsui1982} are observed in two-dimensional electron systems (2DES) under high magnetic fields at low temperatures and are studied intensively. The integer QHE (IQHE) is usually investigated within the single particle pictures, where the system is highly disordered~\cite{Schweitzer1985,Kramer2003}. Under QHE conditions the energy dispersion is quantized due to perpendicular magnetic field $B$ and energy  (Landau) levels are given by $E_{n}=\hbar \omega_{c}(n+1/2)$, where $\omega_{c} = eB/mc$ is the cyclotron frequency and $n$ is the Landau index. In the experimental realization of the IQHE, Hall conductance $\sigma_{H}$ assumes quantized values in units of $\nu e^{2}/h$ where $h$ is Planck's constant and $e$ is the elementary charge, while longitudinal conductance $\sigma_{L}$ vanishes if the filling factor $\nu$ ($=2\pi l_{B}^{2}n_{el}$) is an integer. Here $l_{B}(=\sqrt{\hbar/eB})$ is the magnetic length and $n_{el}$ is the electron density. Once $\nu$ is an integer Landau levels are fully occupied, therefore density of states vanishes at the Fermi energy ($E_F$) and as a consequence electrostatic potential can not be screened, hence electron density is constant. Due to this property these states are called incompressible strips. Screening is nearly perfect if $E_F=E_n$ (i.e. within the compressible strips) and the electron density varies similar to the $B=0$, while potential distribution is approximately flat.

In a 2DES, disorder emanates from inhomogeneous distribution of dopant ions that creates the confinement potential for the electrons~\cite{Schweitzer1985, Kramer2003, Chklovskii1992, Lier1994, Oh1997}. It is important to clarify the range definition, while there is a misunderstanding between the theoreticians and experimentalists: Theoreticians usually refers to the range of single impurity disorder whereas experimentalists take the range of total potential fluctuations. Namely, for theoreticians, short range is at the order of nanometers, in contrast for experimentalists it is at the order of few micrometers.  In this study we analyze the disorder potential considering short range single impurity potentials, however, due to high number of such short range impurities the total potential fluctuations are the long range. The effect of long range single impurity potential fluctuations are usually discussed within the well known classical localization picture~\cite{Fogler1994}, which is beyond the scope of this report.

This work investigates the effect of disorder on local electron temperature under IQHE conditions within the screening theory. We base our disorder arguments on one of our previous works, where the disorder potential is studied considering only the effect of long range potential fluctuations, while the electron temperature is assumed to be uniform and being equal to lattice temperature ~\cite{Gulebaglan2012}. However, in this report we move a step further and include local electron temperature calculations using the theory of thermo-hydrodynamics in linear response regime. In this regime, electron temperature is proportional to the current density which depends on the spatial distribution of compressible and incompressible strips ~\cite{Yurdasan2012, Yurdasan2013}. Thermo-hydrodynamic theory is described by conservation of electron number and thermal flux densities, following Akera and his co-workers ~\cite{Akera2000,Akera2001,Akera2002,Akera2005,Kanamaru2006}.

\section{Model}\label{model}

\subsection{Screening theory}

We assume a two dimensional electron gas (2DEG) to be located in $xy-$ plane with translation invariance in the current ($x-$) direction and lateral confinement in $y-$. Electrons are confined by the background potential $V_{\mathrm{bg}}(y)$ generated by ionized donors, which are assumed to be distributed uniformly in the $xy-$ plane on average. The Hartree potential $V_{\mathrm{H}}(y)$ is determined by the electron density and describes the interaction between electrons approximated by the direct Coulomb potential. The cumulative impurity potential $V_{\mathrm{imp}}(y)$ results from the inhomogeneous distribution of the donors. Thus, the effective potential within the semi-classical (i.e., Thomas-Fermi) approximation can be written as
\begin{equation}\label{1}
V(y)=V_{\mathrm{bg}}(y)+V_{\mathrm{H}}(y)+V_{\mathrm{imp}}(y),
\end{equation}
where,
\begin{equation}\label{2a}
V_{\mathrm{bg}}(y)=-E_{\mathrm{bg}}^{0}\sqrt{1-\left(\frac{y}{d}\right)^{2}}, \quad \quad E_{\mathrm{bg}}^{0}=\frac{2\pi e^{2}}{\kappa}n_{0}d,
\end{equation}
with $E_{\mathrm{bg}}^{0}$ is the strength of the confining potential due to the homogeneous background donor distribution, $n_{0}$ is the average donor density and $2d$ being the sample width ~\cite{Siddiki2006, Siddiki2010}. The Hartree potential generated by the 2DEG can be written as:
\begin{equation}\label{3a}
V_{\mathrm{H}}(y)=\frac{2e^{2}}{\kappa}\int^{d}_{-d}dy^{'}K(y,y^{'})n_{\mathrm{el}}(y^{'}),
\end{equation}
where $\kappa$ is the average background dielectric constant (12.4 for GaAs), and the kernel $K(y,y^{'})$ solves Poisson's equation under the given boundary conditions given by
\begin{equation}\label{4a}
K(y,y^{'})=\ln\left|\frac{\sqrt{(d^{2}-y^{2})(d^{2}-y^{'2})}+d^{2}-y^{'}y}{(y-y^{'})d}\right|.
\end{equation}

The kernel and potentials are calculated assuming that all charges (i.e. electrons and donors) and metallic gates reside on the $xy-$ plane following the seminal work by D. B. Chklovskii et.al. ~\cite{Chklovskii1992}. The same approach is also used when considering an external dissipative current by R. R. Gerhardts and his coworkers~\cite{Guven2003,Siddiki2004}.

Once the background potential and the total number of electrons are fixed by $n_0$ and the chemical potential $\mu_{\mathrm{ec}}$, respectively, it is possible to obtain spatial electron density distribution utilizing a lowest level density functional approach. Assuming that $V(y)$ varies slowly on the scale of magnetic length $l_{B}=\sqrt{\hbar/eB}$, one can calculate the electron density $n_{\mathrm{el}}(y)$ within the Thomas-Fermi approximation via,
\begin{equation}\label{5a}
n_{\mathrm{el}}(y)=\int dE D(E) f(E+V(y)-\mu_{\mathrm{ec}}),
\end{equation}
where $D(E)$ is the density of states and $f(E)$ being the Fermi function~\cite{Guven2003,Siddiki2004,Gerhardts2008}. In local equilibrium, the energy dispersion of an electron is determined by the Fermi function $f(E)=1/[\exp(\varepsilon-\mu_{\mathrm{ec}})/k_{\mathrm{B}}T_{\mathrm{e}}+1]$ with $\varepsilon$  the energy and $T_{\mathrm{e}}$ the electron temperature. 

In the following sections we include the impurity effects by twofold: first, the contribution of single impurity potentials will be added to total confinement potential as long-range modulations and second, impurity potentials will be used to determine the local transport coefficients as described in Ref.~\cite{Siddiki2004} 

\subsection {Many impurities: potential fluctuations}

Extracted from local probe experiments~\cite{Dahlem10:121305,Suddards12:083015} and theoretical works~\cite{SiddikiGerhardts2004,Gerhardts:2017,Gerhardts:2019}, the effective (total) potential fluctuates over a length scale up to two-hundred nanometers for low impurity concentrations and can be as large as micrometers for high impurity concentrations. Here, we investigate the effect of such potential fluctuations within the screening theory by taking into account overall potential variation from an inhomogeneous donor distribution described by a modulation potential;
 $$V_{\rm mod}(y)=V_0\cos{(\frac{ 2 \pi y m_{p}}{2d}}),$$
where $m_p$ is the modulation period conserving the assumed boundary conditions and thereby defines the quality (or mobility) of the sample. Note that, the modulation amplitude $V_{0}$ is independent of the sample length, since it is determined by the crystal growth parameters and conditions. In our calculations we vary this amplitude to model the sample quality by setting $V_{0}/E_{F}^{0}$ to small values for high mobility and relatively large values for the opposite case. 

To summarize: A single donor, which is distributed in a inhomogeneous manner, creates a single impurity potential. The accumulation of such single impurity potentials yield long range fluctuations at the total potential which may vary between few hundred nanometers to few micrometers. In the following we investigate the effect of this potential fluctuations on the formation of incompressible strips and thereby on local electron temperature distribution. In previous works, the effects of screening, external magnetic field, sample properties such as width and geometry were taken into account. However, the effect of disorder on electron temperature was left unresolved.

Here we aim to study this local temperature distribution, considering the effect of the spacer thickness (i.e. the distance between the donors and 2DEG, $z$), assuming that the amplitude of the disorder potential is damped up to 50$\%$ of the Fermi energy, motivated by the experimental findings ~\cite{Dahlem10:121305} and the references given there. 

The amplitude is determined essentially by the damping due to screening effect. One can estimate $V_0$ from a simple relation between the external potential $V_{\mathrm ext}(q)$ and screened potential $V_{\mathrm ext}(q)$ in momentum space $q$ by;
\begin{equation}\label{5a_b}
V_{\mathrm scr}(q,z)=V_{\mathrm ext}(q)e^{-|qz|}/\varepsilon(q), \varepsilon(q)=1+2/ (a_B^*|q|).
\end{equation}
Here, $\varepsilon(q)$ is $q$ dependent dielectric function and  $a_B^*(\sim 9.8$ nm, for GaAs) being the effective Bohr radius. 
Hence we can assume that the low mobility is expressed by the modulation amplitude $V_{0}/E_{F}^{0}=0.5$, while the high mobility is defined by $V_{0} /E_{F}^{0}=0.05$. We tabulated the relation between the mobility and geometrical properties of the sample in Table ~\ref{table:1}. Detailed description can be found in Ref. \cite{Gulebaglan2012}. It is important to note that the mobility of the sample depends on width of the concerned device while number of peaks due to impurity modulations does effect the screening.

\begin{table}[h!]
\begin{tabular}{|c|c|c|c|}
  \hline
  mobility & $m_p$ (10 $\mu$m) & $m_p$ (2 $\mu$m) & $V_0/E_F^0$ \\
  \hline
  \hline
  low & 19-20& 5-6 & 0.5 \\
  \hline
  intermediate 1 & 9-10 & 2-3 & 0.5 \\
  \hline
  intermediate 2 & 19-20 & 5-6& 0.05 \\
  \hline
  high & 9-10 & 2-3 & 0.05 \\
  \hline
\end{tabular}
\centering
\caption{A qualitative comparison of the mobility considering magnetic field also in the presence of the self-consistent screening.}\label{table:1}
\end{table}
In the next subsection we introduce the fundamental equations to describe local electron temperature utilizing the concepts of thermo-hydrodynamic theory following Akera and his co-workers~\cite{Akera2000,Akera2001,Akera2002,Akera2005}.

\subsection{Thermo-hydrodynamic Theory}

In our approach to obtain local electron densities our main approximation relies on the facts that the electron number and the total energy of the system $\epsilon$ are preserved in the system. Hence, We consider two thermo-hydrodynamical equations which conserve the electron number conservation and thermal energy 
\begin{equation}\label{9}
\frac{\partial n_{\mathrm{el}}}{\partial t}=-\nabla.\mathbf{j}_{n_{\mathrm{el}}},
\end{equation}
\begin{equation}\label{10}
\hspace{1cm}\frac{\partial \varepsilon}{\partial t}=-\nabla.\mathbf{j}_{\varepsilon}-P_{\mathrm{L}},
\end{equation}
where $\mathbf{j}_{n_{\mathrm{el}}}$ is the number flux density and $\mathbf{j}_{\varepsilon}$ is the energy flux density. $P_{\mathrm{L}}$ is the energy loss per unit area  due to the heat transfer between electrons and phonons~\cite{Akera2005}. The thermal flux density is given by
\begin{equation}\label{11}
 \mathbf{j}_{q}= \mathbf{j}_{\varepsilon}-\mu_{\mathrm{ec}}\mathbf{j}_{n_{\mathrm{el}}}.
\end{equation}
where $ \mathbf{j}_{\varepsilon}$ is the energy flux density and $\mu_{\mathrm{ec}}$ is the electrochemical potential. As we assume translational invariance in the current direction, the electron temperature $T_{\mathrm{e}}$ is independent of $x$. According to above given boundary conditions, Eqs.~(\ref{9}) and (\ref{10}) become
\begin{equation}\label{12}
\Delta j_{n_{\mathrm{el}}}(y)=0 \quad \quad (-d<y<d),
\end{equation}
\begin{equation}\label{13}
\nabla_{y}(\Delta j_{q}(y))+eE_{x}j_{n_{\mathrm{el}}}(y)+P_{\mathrm{L}} =0,
\end{equation}
with the deviations from the equilibrium values~\cite{Kanamaru2006}. These two equations imply that there is no charge transfer in lateral direction and all the heat dissipation can be embedded to parameter $P_L$, which is taken to be ... in accordance with previous works~\cite{Akera2005,Kanamaru2006}. 

\section{Results and Discussion}

 In our calculations, we choose the Fermi energy at the center $E_{F}^{0}$ as the energy scale and the magnetic length $l_{B}$ as a length scale. $E_{F}^{0}$ is equal to $n_{\mathrm{el}}(0; B=0, T=0)/D_{0}$ regarding to the electron density at the center and $D_{0}=m^*/(\pi\hbar^{2})$ the density of state of the 2DES at $B=0$ with effective electron mass $m^*=0.067m_e$. Here we first fix the average donor density $n_0=3.61\times10^11$ cm$^{-2}$, which yields to an average electron density $\bar{n}_{\mathrm el}\sim 3.0\times10^11$ cm$^{-2}$ in the interval $-d<y<d$. The magnetic length $l_{B}$ is important for understanding quantum mechanical effects, e.g. the width of the wave function determines tunneling probabilities and via $\nu$ describes the occupation of the Landau levels. Recall that the Landau levels are fully occupied at integer $\nu$ and is known as the incompressible state with constant electron density distribution ~\cite{Siddiki2004,SiddikiGerhardts2004}.

\begin{figure}[t]
\begin{center}
\includegraphics[scale=.5]{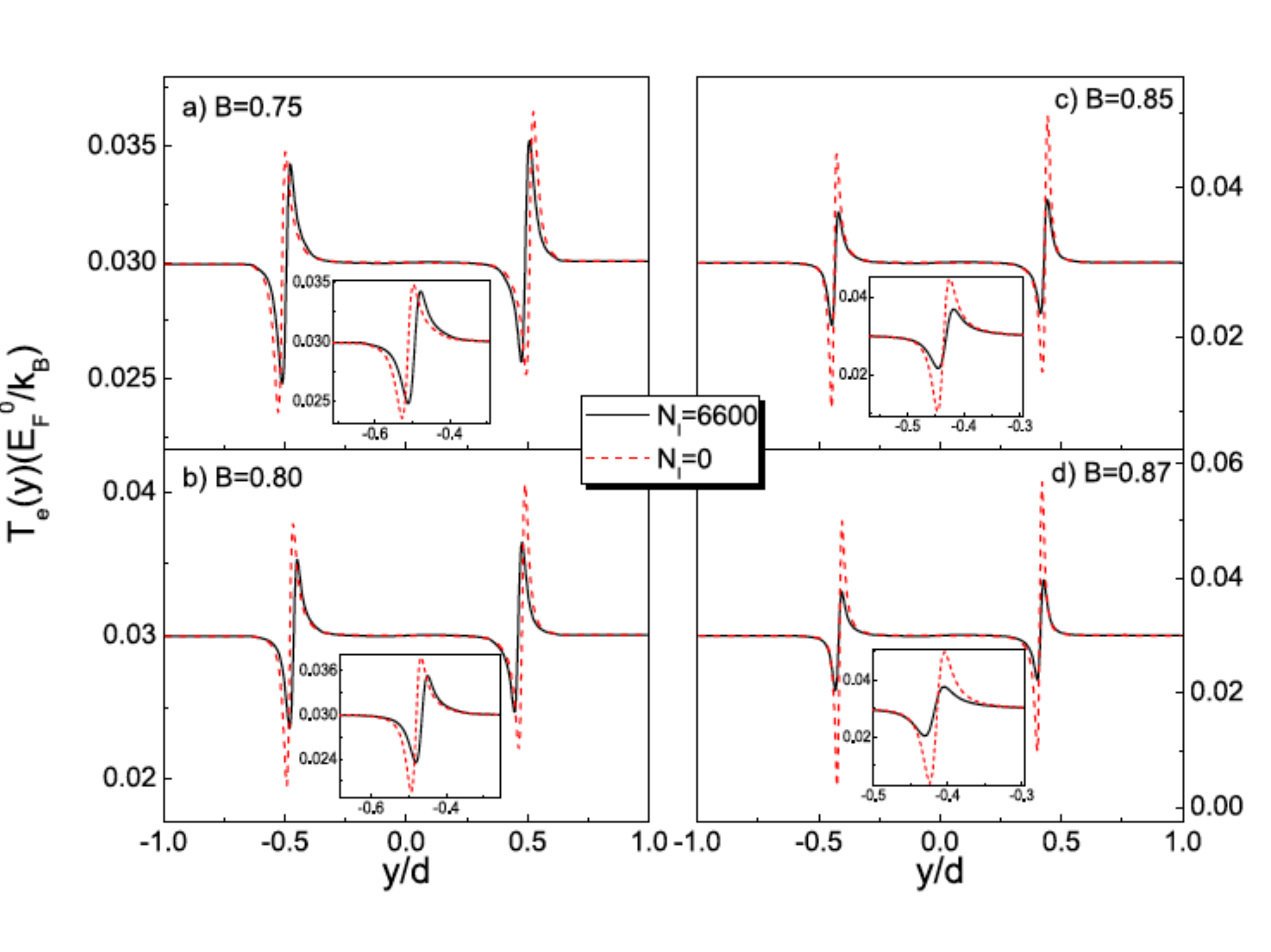}
\caption{ Calculated the electron temperature $T_{\mathrm{e}}$ versus position considering the different values of magnetic fields, $\hbar\omega_{c}/E_{F}^{0}=\Omega_{c}/E_{F}^{0}$=0.75, 0.80, 0.85 and 0.87 for the number of impurity $N_{l}=0$ and $N_{l}=6600$  at the lattice temperature, $T_{\mathrm{L}}=0.03$ $E_{F}^{0}/k_{B}$. Sample parameters are $d=1$ $\mu m$ and $V_{0}=0.5$.  The insets show the enlarged region for the left side. (Fermi energy at the center $E_{F}^{0}=12.68$ meV and donor density $n_{0}=3.61\times 10^{11}$ $cm^{-2}$).}
\label{fig:figure1}
\end{center}
\end{figure}

Given the above calculation scheme and sample parameters we obtained the local electron temperature at different magnetic fields. The results are shown in Fig. 1, considering a $2~\mu$m wide sample with a modulation potential amplitude  $V_{0}=0.5$, at lattice temperature $T_{L}=0.03$ $E_{F}^{0}/k_{B}$ for different impurity concentrations; namely for $N_{l}=0$ (broken (red) lines) and $N_{l}=6600$ (solid (black) lines). At sufficiently low lattice temperatures, we observe that current flows along the incompressible strips in accordance with the literature~\cite{Guven2003,Siddiki2004}. Interestingly one side of the sample heats up, the other side is cooled down~\cite{Yurdasan2012}, which is also consistent with the asymmetric distribution of current reported by experimental findings~\cite{Dahlem10:121305}. At a first glance we see large variations at the local electron temperature in the absence of impurity. The amplitude of variations become smaller when considering impurities, since the kinetic energy of electrons are diminished due to dissipation, resulting from momentum relaxation processes. Once impurities are included to self-consistent calculations the local transport coefficients (i.e. $\sigma_L(y)$ and $\sigma_H(y)$)) lead to higher scattering, hence increased dissipation. Moreover due to broadened DOS, narrower incompressible strips are formed at low mobility samples. Our calculations also show that the variation of electron temperature strongly depends on the magnetic field (cf. insets of Fig.1). Since the narrow edge incompressible strips slide towards the center of the sample while increasing the strength of $B$ they become wider, therefore, scattering increases. These results are in agreement with calculations by Gerhardts and his co-workers ~\cite{Siddiki2004,Gerhardts2008,SiddikiGerhardts2004,Siddiki2007,SiddikiGerhardts2007}.

In Figure 2, the electron temperatures $T_{e}$ are shown at an elevated lattice temperature for different modulation amplitudes considering three characteristic $B$ values. Obtained data shows that the deviation of electron temperature decreases as modulation potential amplitude increases. This is due to the restricted transport of electrons in the system as a result of increased $V_{0}$. In other words, electrons are forced to be confined in narrower incompressible strips due to steeper confinement potential. As a result, the kinetic energy of the electrons is decreased, which suggests decrease in the local electron temperature, similar to the case where number of impurities is increased. Consistent with our previous conclusion, we observe similar behavior depending on magnetic field strength.

\begin{figure}[t]
\begin{center}
\includegraphics[scale=0.5]{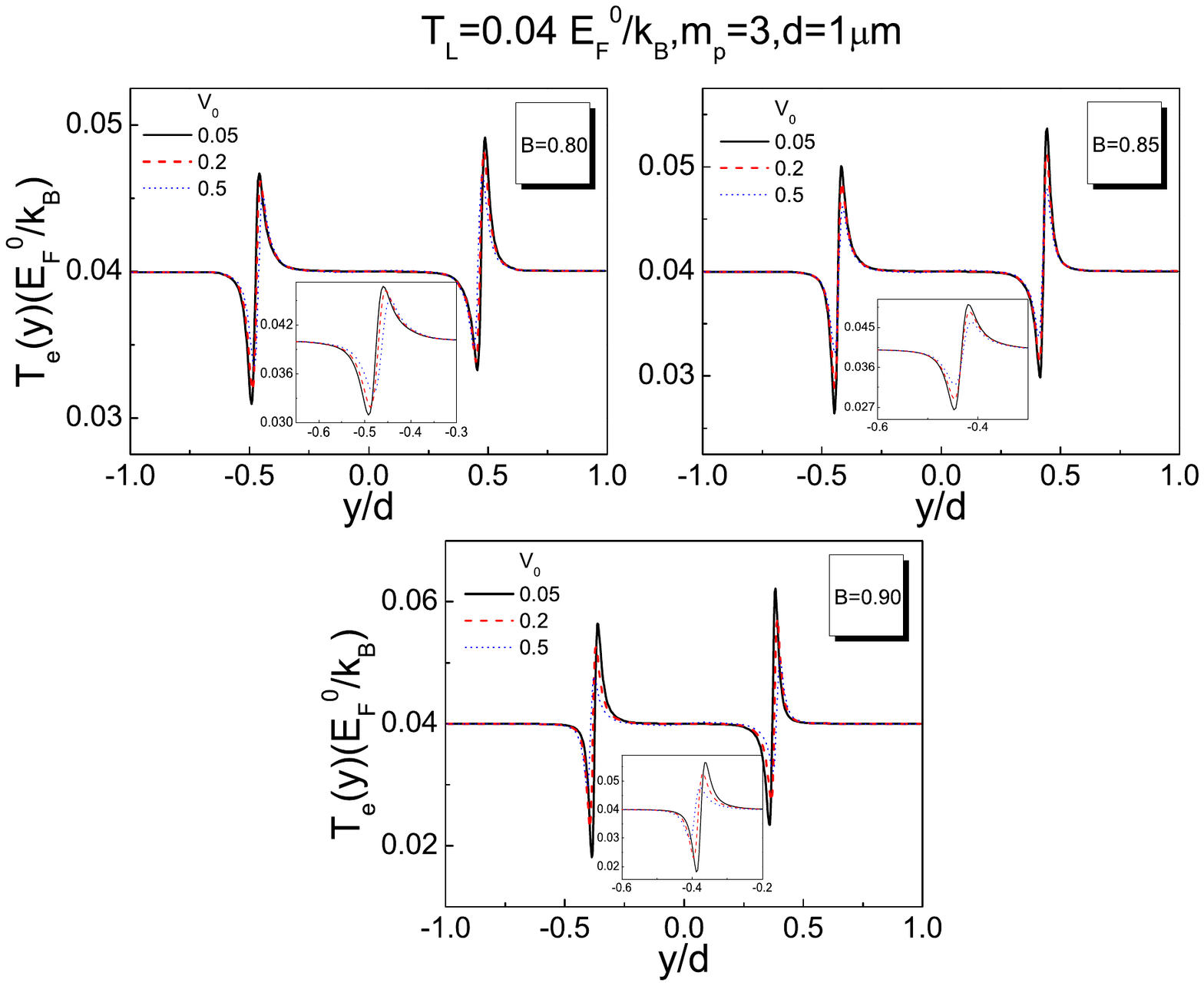}
\vspace{-4cm}
\caption{The electron temperature $T_{e}$ versus position calculated for different values of the modulation potential $V_{0}$. The calculations are done at $T_{\mathrm{L}}=0.04$ $E_{F}^{0}/k_{B}$ lattice temperature considering impurity  $N_{l}=6600$ and repeated for three different values of magnetic field $\hbar\omega_{c}/E_{F}^{0}=0.80, 0.85$ and $0.90$. The insets show the enlarged region for the left side.}
\label{fig:figure2}
\end{center}
\end{figure}

\begin{figure}[t]
\begin{center}
\includegraphics[scale=0.5]{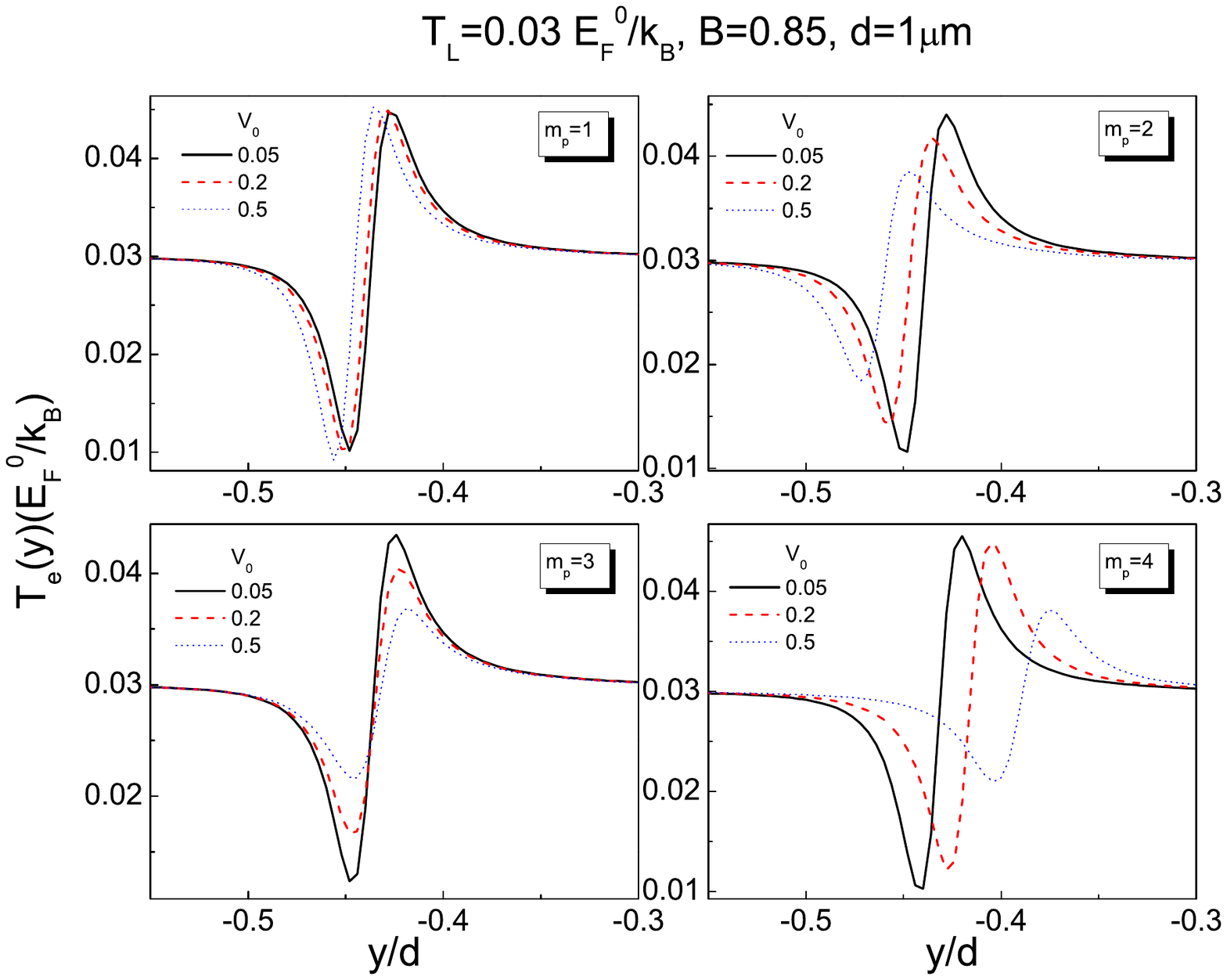}
\vspace{-4cm}
\caption{The electron temperature $T_{e}$ versus position considering the magnetic field $\hbar\omega_{c}/E_{F}^{0}=\Omega_{c}/E_{F}^{0}=0.85$
 with the sample width of $d=1$ $\mu$m at  $T_{L}=0.03$ $E_{F}^{0}/k_{B}$ }
\label{fig:figure3}
\end{center}
\end{figure}

Figure 3 presents local electron temperature distributions for $d=1\mu m$ wide sample at $T_{L}=0.03$ $E_{F}^{0}/k_{B}$ under magnetic field of $\hbar\omega_{c}/E_{F}^{0}=0.85$ for different modulation potentials $V_{0}$ considering various mobilities $m_{p}$. Mobility is described by the scattering processes within the sample, hence low mobility corresponds to high scattering rates. The effects of disorder on the integer quantized Hall system within the screening theory reported by A. Siddiki et. al. ~\cite{Siddiki2006} predict that low mobilities give rise to larger (or wider) incompressible strips, as it is observed in many experiments ~\cite{Horas2008,SiddikiHoras2009,Haug1987}. If modulation potential is set to $V_{0}=0.05$, choosing the range to be $m_{p}=2-3$ represents high mobility sample. Besides, current density in the system is related to the drift velocity of the charges contributing to conductivity. Higher mobility suggests increasing drift velocity of electrons and accordingly larger kinetic energies. This feature implies that the scattering probability per unit time decreases. From this behavior, we can conclude that the local electron temperature increases for higher mobility samples. Obviously, the incompressible strips move towards to the center of the sample faster for increased mobility, which is in agreement with previous report in the literature ~\cite{Gulebaglan2012}.

\section{Conclusion}

In conclusion, we investigate the local electron temperature in the presence of disorder considering a dissipative (integer) quantum Hall system. Our calculations are based on screening and thermo-hydrodynamic theories which are good agreement with the previous investigations. As expected, the deviation of local electron temperature increases/decreases due to a variation of magnetic field and modulation amplitude. Also this work shows that the deviation of local electron temperature, observed in incompressible strips, strongly depends on the number of impurity atoms and mobilities due to the fact that such parameters affect the kinetic energy of electrons.  Therefore, we conclude that the deviation of the local electron temperature and its location strongly depends on the kinetic energy.

\end{document}